\documentclass[fleqn,10pt]{wlscirep}

\usepackage{soul}
\usepackage{longtable}
\usepackage{bm}
\usepackage{dcolumn}
\usepackage{epsfig}
\usepackage{graphicx}
\usepackage{subfigure}
\usepackage{comment}
\usepackage{amsfonts}
\usepackage{amsmath}
\usepackage{amssymb}
\usepackage{times}
\usepackage{appendix}
\usepackage{color}

\usepackage{epstopdf}
\usepackage{epsfig}

\newcommand{\be}{\begin{equation}}
\newcommand{\ee}{\end{equation}}
\newcommand{\bea}{\begin{eqnarray}}
\newcommand{\eea}{\end{eqnarray}}

\begin{document}

\title{A Comment on the One Step Quantum Key Distribution Based on EPR Entanglement}

\author[*]{Anirban
Pathak}
\author[$\dag$]{Kishore Thapliyal}
\affil{Jaypee Institute of Information Technology, A-10, Sector-62,
Noida, UP-201307, India}
\affil[*]{email: anirban.pathak@gmail.com}
\affil[$\dag$]{email: tkishore36@yahoo.com}

\begin{abstract}
Recently, in Sci. Rep. \textbf{6} (2016) 28767, Li et al., have proposed
a scheme for quantum key distribution using Bell states. This comment
provides a proof that the proposed scheme of Li et al., is insecure
as it involves leakage of information. Further, it is also shown that
all the error rates computed in the Li et al.'s paper are incorrect
as the authors failed to recognize the fact that any eavesdropping
effort will lead to entanglement swapping. Finally, it is established
that Li et al.'s scheme can be viewed as an incrementally (but incorrectly)
modified version of the existing schemes based on Goldenberg Vaidman (GV) subroutine. 
\end{abstract}

\date{\today}
\maketitle

Recently, Li et al. \cite{Li-2016} have proposed a one step scheme
for quantum key distribution (QKD) using a Bell states. The scheme
is not secure and the claims made by the authors unfortunately appear
to be inaccurate for various reasons as elaborated in this comment.
Firstly, they claimed their scheme to be novel and tried to support
their claim by stating ``Compared with the \textquotedblleft Pingpong\textquotedblright{}
protocol involving two steps, the proposed protocol does not need
to store the qubit and only involves one step''. This is a comparison
of an apple with an orange, as the Pingpong protocol is designed for
quantum secure direct communication (QSDC), where information can
be transmitted using quantum resources without the prior generation
of keys. Naturally, a scheme of QKD should not be compared with a
scheme of QSDC. To visualize the naiveness of their claim and part
of the exercises followed in their paper, one may note that even the
well known BB84 \cite{bb84} and B92 \cite{b92} protocols for QKD
do not require quantum memory, if one wishes to design BB84 type scheme
using EPR states, Alice can simply prepare some copies of Bell states
(say $|\phi^{+}\rangle=\frac{1}{\sqrt{2}}\left(|00\rangle+|11\rangle\right)=\frac{1}{\sqrt{2}}\left(|++\rangle+|--\rangle\right)$,
and send the second qubits of all the Bell pairs to Bob, who subsequently
randomly measures his qubit in $X$ basis or $Z$ basis. Alice also
performs the same measurement and they keep all those cases where
the measurement basis used by them are the same. For eavesdropping
check they may compare half the results, if no signature for eavesdropping
is found their key will be exactly correlated. A scheme equivalent
to this scheme, which is one step and uses EPR state, is discussed
by some of the present authors as Protocol 2 in \cite{my-book}, where
to increase the efficiency, Bob used to announce his basis and Alice
used to subsequently measure in the same basis. However, such a scheme
is neither efficient (as it uses costly quantum resource, like entanglement,
which is not required) nor novel (as the scheme is equivalent to BB84).
The one step EPR-based scheme proposed by Li et al., is slightly different
from this simple minded EPR-based one step scheme of QKD described
above as in this case Bob performs Bell measurement. In what follows
we would show that security of such a scheme appears from entanglement
swapping. Unfortunately, Li et al., missed this point and analyzed
the security of their protocol in analogy with BB84 scheme, and naturally
such an incorrect analysis led to incorrect values of error rates.
Before, we continue with our comments and analysis, for the sake of
completeness of this comment, we would like to briefly describe Li
et al.'s protocol.

\section*{Analysis of Li et al.'s protocol}

Let us first summarize the first QKD protocol of Li et al., which
requires memory. In fact, the protocol is nothing but a DSQC protocol
proposed in the recent past by some of the present authors \cite{With-chitra-IJQI},
where Alice is sending random bits instead of a meaningful message.
Specifically, Alice prepares a Bell state corresponding to each 2-bits
of her key, i.e., $|\phi^{+}\rangle,\,|\phi^{-}\rangle,\,|\psi^{+}\rangle$
and $|\psi^{-}\rangle$ for sending 00, 01, 10, 11, respectively.
This can also be viewed as Alice preparing $\frac{N}{2}$ copies of
$|\phi^{+}\rangle$ and applying Pauli operations $I,\,Z,\,X$ and
$iY$ to encode 00, 01, 10, 11, respectively. Subsequently, she permutes
the string of encoded particles and sends it to Bob, who uses memory
to store the qubits other than decoy qubits. They proceed for final
key generation only if the channel is ensured safe. It is worth noting
here that a QKD protocol can be modified to a direct communication
protocol by incorporating memory, whereas a direct communication protocol
can always be converted to a QKD protocol if the sender sends random
key rather sending a meaningful message. 

The second protocol is a modification of the first protocol to circumvent
the use of memory. Specifically, the authors tried to design an entangled-state-based
protocol in analogy with BB84 protocol, where Alice prepares two Bell
states for each 4-bits of her secret key and probabilistically swap
the second particle of the first Bell state with the first particle
of the second Bell state (or in other words, arrange the 4 particles
in one of the following orders $\left\{ \left(1,2\right),\left(3,4\right)\right\} $
or $\left\{ \left(1,3\right),\left(2,4\right)\right\} ,$ where the
particles $\left(1,2\right)$ and $\left(3,4\right)$ are entangled
with each other). This probabilistic swap is equivalent to a restricted
randomization process that does not allow all possible permutations,
for example, it does not allow $\left\{ \left(1,4\right),\left(2,3\right)\right\} $
as a valid sequence. Thus, it leaves only two choices for Bob, i.e.,
either to measure the string as such ($\left\{ \left(1,2\right),\left(3,4\right)\right\} $)
or after swapping second and third qubits in the string ($\left\{ \left(1,3\right),\left(2,4\right)\right\} $).
After repeating this procedure for all the key bits, Alice and Bob
first check some of the decoy qubits, and if errors are low they proceed
to key generation.

In both the schemes proposed in Ref. \cite{Li-2016} and summarized
here, the authors claimed that the security arises due to Heisenberg's
uncertainty principle and no-cloning. However, it should be noted
that the security in a cryptographic protocol is achieved using splitting
the useful information in 2 or more pieces, in such a way that unavailability
of all these pieces makes it impossible to extract the secret \cite{With-chitra-IJQI}.
For example, in BB84 protocol, Alice sends the quantum piece (qubits)
to Bob and withholds classical information regarding choice of basis
for each qubit. Similarly, in the present protocols, Alice withholds
the classical information of order of permutation. This principle
of origin of security due to temporal or spatial separation of entangled
particles causing ignorance for an eavesdropper has been discussed
in the past as GV subroutine in Refs. \cite{With-chitra-IJQI,QKA-cs,With-preeti,dsqc-ent-swap}.
In GV subroutine, if Eve incorrectly chooses particles from two entangled
states to measure in the basis they were initially prepared, then
it causes entanglement swapping, which leaves detectable traces for
the receiver. Interestingly, its equivalence with BB84 subroutine,
in which single qubits are used as decoy qubit, has also been addressed
over noisy channels in the recent past \cite{RDshrama2015}. In GV
subroutine and any other scheme, where particle order of entangled
particles are permuted before transmitting through the channel, eavesdropping
efforts lead to entanglement swapping and thus a detectable trace
of eavesdropping. 

Interestingly, the authors completely missed the role of entanglement
swapping, which is reflected in the erroneous computation of error
rates and mutual information. To visualize this point, let us look
in Group 2 in Table 3 of Ref. \cite{Li-2016}; if Eve selects the
wrong pair of qubits to measure in the Bell basis that would have
led to entanglement swapping. Specifically, due to entanglement swapping,
the initial choice of states $|\phi^{+}\rangle_{12}|\psi^{+}\rangle_{34}$
would become 
\begin{equation}
\begin{array}{lcl}
|\phi^{+}\rangle_{12}|\psi^{+}\rangle_{34} & = & \frac{1}{2}\left(|\phi^{+}\rangle_{13}|\psi^{+}\rangle_{24}+|\phi^{-}\rangle_{13}|\psi^{-}\rangle_{24}+|\psi^{+}\rangle_{13}|\phi^{+}\rangle_{24}+|\psi^{-}\rangle_{13}|\phi^{-}\rangle_{24}\right),\end{array}\label{eq:phiPl-psiPl}
\end{equation}
which will never lead to a measurement outcome by Eve $|\psi^{+}\rangle_{13}|\psi^{-}\rangle_{24}$.
Similarly, if Bob performs a measurement after Eve's measurement using
an incorrect sequence that would never lead to $|\psi^{+}\rangle_{12}|\phi^{-}\rangle_{34}$.
To visualize this consider that Eve's measurement yields $|\phi^{+}\rangle_{13}|\psi^{+}\rangle_{24}=\frac{1}{2}\left(|\phi^{+}\rangle_{12}|\psi^{+}\rangle_{34}+|\phi^{-}\rangle_{12}|\psi^{-}\rangle_{34}+|\psi^{+}\rangle_{12}|\phi^{+}\rangle_{34}+|\psi^{-}\rangle_{12}|\phi^{-}\rangle_{34}\right)$,
then a subsequent measurement of Bob in a correct sequence would yield
one of the following 4 possible results with equal probability: $|\phi^{+}\rangle_{12}|\psi^{+}\rangle_{34},\,|\phi^{-}\rangle_{12}|\psi^{-}\rangle_{34},\,|\psi^{+}\rangle_{12}|\phi^{+}\rangle_{34},\,|\psi^{-}\rangle_{12}|\phi^{-}\rangle_{34}.$
Clearly, in this particular case, Bob cannot obtain $|\psi^{+}\rangle_{12}|\psi^{-}\rangle_{34}$.
Similarly, we can expand the other 3 possible outcomes of Eve's measurement
as 
\begin{equation}
\begin{array}{lcl}
|\phi^{-}\rangle_{13}|\psi^{-}\rangle_{24} & = & \frac{1}{2}\left(|\phi^{+}\rangle_{12}|\psi^{+}\rangle_{34}+|\phi^{-}\rangle_{12}|\psi^{-}\rangle_{34}-|\psi^{+}\rangle_{12}|\phi^{+}\rangle_{34}-|\psi^{-}\rangle_{12}|\phi^{-}\rangle_{34}\right),\end{array}\label{eq:phiM-psiM}
\end{equation}
\begin{equation}
\begin{array}{lcl}
|\psi^{+}\rangle_{13}|\phi^{+}\rangle_{24} & = & \frac{1}{2}\left(|\phi^{+}\rangle_{12}|\psi^{+}\rangle_{34}-|\phi^{-}\rangle_{12}|\psi^{-}\rangle_{34}+|\psi^{+}\rangle_{12}|\phi^{+}\rangle_{34}-|\psi^{-}\rangle_{12}|\phi^{-}\rangle_{34}\right)\end{array},\label{eq:psiPl-phiPl}
\end{equation}
and 
\begin{equation}
\begin{array}{lcl}
|\psi^{-}\rangle_{13}|\phi^{-}\rangle_{24} & = & \frac{1}{2}\left(|\phi^{+}\rangle_{12}|\psi^{+}\rangle_{34}-|\phi^{-}\rangle_{12}|\psi^{-}\rangle_{34}-|\psi^{+}\rangle_{12}|\phi^{+}\rangle_{34}+|\psi^{-}\rangle_{12}|\phi^{-}\rangle_{34}\right).\end{array}\label{eq:psiM-phiM}
\end{equation}
Now, we can easily see that whatever be the measurement outcome of
Eve, for an initial state $|\phi^{+}\rangle_{12}|\psi^{+}\rangle_{34}$,
Eve can never obtain $|\psi^{+}\rangle_{12}|\phi^{-}\rangle_{34}.$
A similar mistake can also be observed in Group 3 of Table 4 of Ref.
\cite{Li-2016} for Bob's measurement. 

From the above example, we can easily observe the following from the
perspective of an eavesdropper. Half of the time she chooses the correct
order of particles, and consequently obtains the right bit values,
without leaving any detectable trace of her attack. However, in the
remaining half of the cases (of the groups she intends to attack),
she chooses the wrong order of particles, her measurement causes entanglement
swapping (as discussed through Eqs. (\ref{eq:phiPl-psiPl})-(\ref{eq:psiM-phiM})).
It is worth noting here, still 1/4 of the times she obtains the correct
result, while she obtains wrong results on the remaining 3/4 cases.
Therefore, she obtains the right bit values with probability $\frac{1}{2}\left(1+\frac{1}{4}\right)=\frac{5}{8}$.
For instance, suppose Alice initially prepares $|\phi^{+}\rangle|\psi^{+}\rangle$
as discussed above and sends it to Bob, but the transmitted state
is intercepted by Eve and measured in the Bell basis. Half of the
time, when she chooses the correct particle order she obtains $|\phi^{+}\rangle|\psi^{+}\rangle$,
whereas in the remaining half of the cases her measurement would cause
entanglement swapping, but still Eve obtains the correct result (i.e.,
$|\phi^{+}\rangle|\psi^{+}\rangle$) with $\frac{1}{4}$ probability.
Thus, in half of all the cases, Bob obtains incorrect result (i.e.,
traces of eavesdropping) with $\frac{3}{4}$ probability. Therefore,
an eavesdropping attack is detected with probability $\frac{1}{2}\times\frac{3}{4}=\frac{3}{8}$.
Similar analysis would lead to the same result for other initial states,
too, and we can easily observe that the value for the probability
that Bob succeeds in detecting an eavesdropping attack (correct value
$\frac{3}{8}$ as shown above) for a specific block of 4 qubits, which
has been attacked by Eve is incorrectly computed in Ref. \cite{Li-2016},
where they wrote, `` ... if Bob measures this intercepted Bell states
with the same location Alice sent he gets a random Bell state result,
i.e., an incorrect result with probability of $\frac{15}{16}=93.75\%$.''
As $\frac{15}{16}\gg\frac{3}{8},$ we can conclude that the values
computed in Li et al.'s paper are incorrect and exaggerated. 

Now to illustrate the leakage involved in this protocol, let us look
at the right hand sides of Eqs. (\ref{eq:phiPl-psiPl})-(\ref{eq:psiM-phiM}),
where we can see that in each case, Eve obtains the correct result,
i.e., the state prepared by Alice with probability $\frac{5}{8}$,
while one of the 3 incorrect outcomes with probability $\frac{1}{8}$
each. It is noteworthy that out of the 4 bits of classical information
that Alice sends to Bob, most of it is leaked to Eve as her ignorance
is only $-\frac{5}{8}\log_{2}\frac{5}{8}-3\times\frac{1}{8}\log_{2}\frac{1}{8}=1.54879$
bits. Further, we note that in any realistic attack strategy, Eve
only attacks a fraction $f$ of the transmitted qubits to reduce the
chances of being detected. In the present case, we may assume that
Eve attacks $m$ of the $n$ blocks of transmitted qubits, where each
block contains 4 qubits. Thus, $f=\frac{m}{n}.$ Consequently, average
information extracted by Eve for the fraction $f$ of the transmitted
blocks of qubits attacked by her is 
\[
I\left(A:E\right)=\frac{5f}{8}.
\]

Similarly, the error due to Eve's measurement, that is detected by
Alice and Bob through a comparison is $\frac{3f}{8}$. Therefore,
the mutual information between Alice and Bob is 
\[
I\left(A:B\right)=1-H\left[\frac{3f}{8}\right].
\]
Here, $H\left[u\right]$ corresponds to the Shannon binary entropy.
A protocol qualifies as a secure communication protocol only if $I\left(A:B\right)\geq I\left(A:E\right)$
which is ensured for $f= 0.493875\cong49.39\%$. Thus, tolerable error
rate is $e_{{\rm max}}=\frac{3}{8}\times49.39=18.52\%$. Incidentally,
Li et al.'s incorrect analysis led to a similar value ($11\%$).
However, it would be apt to note that the $e_{{\rm max}}$ reported
here is valid only for a specific eavesdropping attack (measurement-resend
attack), whereas for BB84 protocol, the same is known tightly for
an arbitrary attack and the tolerable error limit against an arbitrary
attack in BB84 is $11\%$ \cite{Shor}. Further, we would like to note that under measurement
resend attack, often higher values of $e_{{\rm max}}$ is found. For example, for Goldenberg Vaidman protocol \cite{GV}
$e_{{\rm max}}$ was computed to be $0.26=26\%$ \cite{With-preeti}. Finally, we would like to conclude
this comment by noting that there is a class of quantum cryptographic
schemes where particle order permutation (PoP) technique is used (see
\cite{QKA-cs,With-preeti,crypt-switch,cdsqc,PoPEx,PoPEx2} and references
therein). This technique was introduced by Deng et al. \cite{PoP}
and is frequently used by us and others \cite{QKA-cs,With-preeti,crypt-switch,cdsqc,PoPEx,PoPEx2}.
In the set of all PoP based schemes, there is a subset of protocols,
in which sender(s) initially create $\frac{N}{2}$ copies of a Bell
state and after encoding her information (if needed) sends all the
particles to Bob, but applies a permutation operator $\Pi_{N}$ before
sending the sequence. Some of these Bell states are used as decoy
qubits. Now, if Eve chooses, two qubits from the sequence that is
transmitted through the channel the probability that the selected
qubit is entangled (i.e., the probability that he performs a measurement
in correct order) is only $\frac{2}{^{N}C_{2}}=\frac{4}{N(N-1)},$
and in that case $I(A:E)=\frac{4}{N(N-1)}\times f\times\frac{5}{4}=\frac{5f}{N(N-1)},$
which reduces exponentially with $N$ and approaches zero for large
$N$. As $I(A:E)$ reduces, $I(A:B)$ increases and consequently $e_{{\rm max}}$
increases. This is in the heart of GV subroutine and PoP based schemes
that uses Bell state (see our earlier works \cite{QKA-cs,crypt-switch,cdsqc}
and references therein). Thus, in Ref. \cite{Li-2016}, an effort
has been made to slightly modify the existing PoP based schemes that
use Bell state, without citing any of them. However, the modification
led to leakage and a lower tolerable error rate under measurement-resend
attack. Further, the main results of \cite{Li-2016} are wrong as
the authors failed to recognize the role of entanglement swapping
in their scheme. 
\\[0.25cm]

 \ \ \
 
 \ \ \

\noindent
{\bf Acknowledgements.}
AP and KT thank Defense Research \& Development
Organization (DRDO), India for the support provided through the project
number ERIP/ER/1403163/M/01/1603.
\\[0.15cm]

\noindent
{\bf Contributions.}
AP conceptualized the problem and has written most part of the manuscript. KT has performed the calculations and written some parts of the manuscript. 
\\[0.15cm]

\noindent
{\bf Competing interests.}
The authors declare no competing financial interests.
\\[0.15cm]

\noindent
{\bf Corresponding author.}
Correspondence to Anirban Pathak (email: anirban.pathak@gmail.com).
\\[0.25cm]


\begin{thebibliography}{10}
\bibitem{Li-2016}Li, J., Li, N., Li, L.L. \& Wang, T. One Step Quantum
Key Distribution Based on EPR Entanglement. \emph{Sci. Rep. }\textbf{6},
28767 (2016).

\bibitem{bb84}Bennett, C. H. \& Brassard, G. Quantum cryptography:
public key distribution and coin tossing. In: Proceedings of the IEEE
International Conference on Computers, Systems, and Signal Processing,
Bangalore, India, pp. 175--179 (1984).

\bibitem{b92}Bennett, C. H. Quantum cryptography using any two nonorthogonal
states. \emph{Phys. Rev. Lett. }\textbf{68}, 3121 (1992).

\bibitem{my-book}Pathak, A. Elements of Quantum Computation and Quantum
Communication\emph{.} CRC Press, Boca Raton, USA (2013).

\bibitem{With-chitra-IJQI}Shukla, C., Pathak, A. \& Srikanth, R.
Beyond the Goldenberg-Vaidman protocol: secure and efficient quantum
communication using arbitrary, orthogonal, multi-particle quantum
states. \emph{Int. J. Quantum Inf.} \textbf{10}, 1241009 (2012).

\bibitem{QKA-cs}Shukla, C., Alam, N. \& Pathak, A. Protocols of quantum
key agreement solely using Bell states and Bell measurement. \emph{Quantum
Inf. Process.} \textbf{13}, 2391--2405 (2014).

\bibitem{With-preeti}Yadav, P., Srikanth, R. \& Pathak, A. Two-step
orthogonal-state-based protocol of quantum secure direct communication
with the help of order-rearrangement technique. \emph{Quantum Inf.
Process.} \textbf{13}, 2731--2743 (2014).

\bibitem{dsqc-ent-swap}Shukla, C. \& Pathak, A. Orthogonal-state-based
deterministic secure quantum communication without actual transmission
of the message qubits. \emph{Quantum Inf. Process. }\textbf{13}, 2099--2113
(2014).

\bibitem{RDshrama2015}Sharma, R. D., Thapliyal, K., Pathak, A., Pan,
A. K. \& De, A. Which verification qubits perform best for secure
communication in noisy channel? \emph{Quantum Inf. Process.} \textbf{15},
1703--1718 (2015).

\bibitem{Shor}Shor, P. W. \& Preskill, J. Simple proof of security
of the BB84 quantum key distribution protocol. \emph{Phys. Rev. Lett.
}\textbf{85}, 441 (2000). 


\bibitem{GV}Goldenberg, L. \& Vaidman, L. Quantum cryptography based on orthogonal states. \emph{Phys. Rev. Lett.} \textbf{75},
1239 (1995).

\bibitem{crypt-switch}Thapliyal, K. \& Pathak, A. Applications of
quantum cryptographic switch: Various tasks related to controlled
quantum communication can be performed using Bell states and permutation
of particles. \emph{Quantum Inf. Process. }\textbf{14}, 2599--2616
(2015).

\bibitem{cdsqc}Pathak, A. Efficient protocols for unidirectional
and bidirectional controlled deterministic secure quantum communication:
Different alternative approaches. \emph{Quantum Inf. Process. }\textbf{14},
2195--2210 (2015).

\bibitem{PoPEx}Wang, J., Zhang, Q. \& Tang, C. J. Quantum secure
direct communication based on order rearrangement of single photons.
\emph{Phys. Lett. A }\textbf{358}, 256--258 (2006). 

\bibitem{PoPEx2}Xiao, L., Long, G. L., Deng, F. G. \& Pan, J. W.
Efficient multiparty quantum-secret-sharing schemes. \emph{Phys. Rev.
A }\textbf{69}, 052307 (2004). 

\bibitem{PoP}Deng, F.-G. \& Long, G. L. Controlled order rearrangement
encryption for quantum key distribution. \emph{Phys. Rev. A} \textbf{68},
042315 (2003).

\end{thebibliography}
\end{document}